# Sixteen-state magnetic memory based on the extraordinary Hall effect.

A. Segal, M. Karpovski, and A. Gerber[*]

Raymond and Beverly Sackler School of Physics and Astronomy,
Tel Aviv University,
Ramat Aviv 69978, Tel Aviv, Israel

We report on a proof-of-concept study of split-cell magnetic storage in which multi-bit magnetic memory cells are composed of several multilevel ferromagnetic dots with perpendicular magnetic anisotropy. Extraordinary Hall effect is used for reading the data. Feasibility of the approach is supported by realization of four-, eight- and sixteen- state cells.

Over the past few years significant efforts have been devoted to development of techniques that further increase the areal density of information in magnetic hard disk drives and magnetic random access memories (MRAM). One of the options consists of using bit patterned media rather than polycrystalline granular media, i.e. media which consists of physically separated magnetic dots, each dot carrying one bit of information [1]. Achieving Tb/in$^2$ recording densities will require a challenging fabrication of sub-10 nm discrete magnetic islands covering a full disk with tight spacing and narrow switching field distribution. Another approach to increasing storage density without reducing the bit size is by increasing the number of bits stored within each dot in a multilevel storage [2]. Two bits per dot have been shown by Albrecht et al. [3] in patterned media comprising stacks of two out-of-plane magnetized films, and by Moritz et al. [4] in media combining in-plane and perpendicular to plane magnetized thin films.

In similar efforts to further increase the ultimate storage density of MRAMs, several multi-state structures and storage schemes have been proposed using both in-plane and perpendicular anisotropy materials. Angular dependent four-state tunneling magnetoresistance cells were proposed by Uemura et al. [5] by using cubic anisotropy of the single crystalline CoFe electrodes; four-state dual spin valve GMR storage was proposed by Law et al. [6] based on Co/Pd and CoFe/Pd multilayers with perpendicular anisotropy; four-state single layer Fe film device was proposed by Yoo et al. [7] combining two out-of-plane states measured by the extraordinary Hall effect (EHE) and two in-plane states detected by the planar Hall effect, and four states in two-layers cell measured by EHE was proposed by Ravelosona and Terris [8].

In this paper we propose an alternative multi-bit storage system in which each memory cell is split among several multilevel dots, and the extraordinary Hall effect (EHE) is used to extract the stored information. Feasibility of the approach is supported by realization of four-, eight- and sixteen- memory state cells.

Fig.1 presents a schematic view of a memory cell comprising two electrically connected ferromagnetic dots $F_1$ and $F_2$. Each dot contains one or several layers exhibiting perpendicular magnetic anisotropy and serving as storage layers. In order to achieve multiple independent magnetic states the storage layers need to be magnetically decoupled and their switching fields have to be well separated so that by following a prescribed sequence of field pulses each layer can be addressed. Information stored in the cell is obtained by using the extraordinary Hall effect (EHE).

In uniform ferromagnetic films the Hall voltage $V_H$ can be expressed as [9]:

$$V_H = \frac{I}{t}(R_o B_z + \mu_0 R_{EHE} M_z) \tag{1}$$

where $I$ is electrical current, $t$ the sample thickness, $R_o$ and $R_{EHE}$ are the ordinary and extraordinary Hall effect coefficients, and $B_z$ and $M_z$ are magnetic field induction and magnetization components normal to the film respectively. In most ferromagnets the EHE component dominates over the ordinary Hall component, and $V_H$ is proportional to the film magnetization (the ordinary Hall component will be neglected in the rest of the discussion). Given a ferromagnetic film with perpendicular magnetic anisotropy exhibiting hysteresis in its magnetization response to applied magnetic field, the remnant out-of-plane magnetization can be detected by measuring a transverse voltage generated due to the EHE, thus reading the magnetic state of the film. A film containing $n$ parallel layers of comparable resistivity, exhibiting distinct EHE resistances and respective switching fields will generate a total EHE voltage of:

$$V_{EHE} = \frac{\mu_0}{n} \sum_i \frac{I_i R_{EHE,i} M_{z,i}}{t_i} \tag{2}$$

where $I_i$ is electric current flowing along the layer $i$, $R_{EHE,i}$, $M_{z,i}$ and $t_i$ are the EHE coefficient, the magnetization normal to the plane and the thickness of the respective layer $i$.

The voltage between points F and C of two electrically connected magnetic films, sketched in Fig.1, is given by:

$$V_{AD,FC} = V_l + \frac{V_{AD,FB} + V_{AD,EC}}{2} = V_l + \frac{\mu_0}{2} \sum_i \frac{I_i R_{EHE,i} M_{z,i}}{t_i n_i} = V_l + V_{EHE,FC} \tag{3}$$

where $V_{AD,FC}$, $V_{AD,FB}$ and $V_{AD,EC}$ are voltages developed between points FC, FB and EC respectively with current flowing between points A and D, $V_l$ is voltage contributed by a longitudinal resistance of the current-carrying portion of two dots and the electrical connection between them, and $V_{EHE,FC}$ is the EHE term between points F and C. Index $i$ includes all parallel sections of films $F_1$ and $F_2$ together, and $n_i$ is the number of layers in the film to which layer $i$ belongs. If the first film is said to contain $n_1$ distinct layers and the second film $n_2$ layers, the number of different combinations of magnetization and therefore also the number of different output voltage signals is $2^{n_1+n_2}$. Since each voltage signal corresponds to a respective memory state, the number of memory states per cell is $2^{n_1+n_2}$.

In regular ferromagnetic materials like Co, Fe, Ni and their alloys the extraordinary Hall resistivity is about two orders of magnitude smaller than the longitudinal resistivity. Therefore, the longitudinal voltage $V_l$ may be much larger than the variable EHE term, resulting in a small relative change of the output signal when switching from one memory state to another. Suppression of the background longitudinal voltage may be achieved by application of the reverse magnetic field reciprocity (RMFR) theorem [10,11]. According to the RMFR: $V_{ab,cd}(H) = V_{cd,ab}(-H)$, where a,b,c and d are four arbitrary locations in a system, the first pair indicates the current leads and the second the voltage leads. In ferromagnetic materials magnetization replaces the applied magnetic field, giving in our case: $V_{AD,FC}(M) = V_{FC,AD}(-M)$. The odd in magnetization EHE term in Eq.3 can be separated from the even $V_l$ term and obtained from two measurements of $V_{AD,FC}$ and $V_{FC,AD}$ made at a given field as:

$$V_{EHE,FC} = \frac{1}{2}(V_{AD,FC} - V_{FC,AD}) \quad (4)$$

where $V_{FC,AD}$ is the voltage measured between points AD when current is flowing between F and C. In a similar way, when using contacts F and E:

$$V_{EHE,FE} = \frac{1}{2}(V_{AD,FE} - V_{FE,AD}) \quad (5)$$

Use of the RMFR protocol allows extraction of the odd in magnetization EHE signal at any given state with no need in achieving an antisymmetric state with reversed magnetizations.

To illustrate the feasibility of the magnetic memory principle described above we present an experimental implementation of four-state, eight-state and sixteen-state cells by using several multilayered Co/Pd films. Fig. 2 presents a realization of a four-state cell. The cell is composed of two samples fabricated by e-beam deposition of Co/Pd multi-strata structures composed of 0.2 nm thick Co and 0.9 nm thick Pd strata repeated six times [Co 0.2 / Pd 0.9]$_6$. The films were deposited on top of 10 nm and 5 nm thick Pd seed layers (samples 1 and 2 respectively), on a room temperature GaAs substrate. The EHE voltages measured across the two samples separately are shown in Fig. 2a as a function of field applied normal to the film's plane. Arrows mark the field sweep direction. Both samples demonstrate sharp and square hysteresis loops, while sample 1, deposited on 10 nm thick Pd and marked by solid circles, exhibits a higher switching field ($H_{s1}$) than that of sample 2 ($H_{s2}$) and a lower EHE signal. Fig. 2b presents the EHE component of the inter-dot voltage obtained by measuring $V_{AD,FE}$ and $V_{FE,AD}$ and using Eq. 5. Solid circles indicate the major hysteresis loop, at which the field was swept between -0.3 T to +0.3 T and back. Open circles indicate two minor loops at which the field sweep was reversed at "writing" field $H_w = 0.1$T ($H_{s2} < H_w < H_{s1}$) and $-H_w$ in states when magnetization vectors in two films are antiparallel. Four distinct states at zero field corresponding to four possible magnetization states in two films: up-up, up-down, down-up and down-down are obtained by following the respective field sweep sequences. Starting from the "reset" up-up state (denoted as state 1) in which both dots are magnetized upward, states 2-4 are achieved by a sequence of additional field pulses: down-down state is achieved by a pulse of -0.3T, state up-down is obtained by a pulse of $-H_w$, and state down-up is reached by applying a pulse of -0.3T, followed by a pulse of $+H_w$.

Importance of the RMFR protocol can be appreciated from Fig. 2c presenting the as-measured total voltage developed between two films during a series of successive field sweeps. High background voltage and resistance drift due to unstabilized temperature impede an accurate cell reading by a direct measurement of the inter-dot signal. In contrast, application of the RMFR protocol provides a clean stable Hall signal shown in Fig.2b.

Cells with more than four states can be obtained by using multilevel films possessing at least two distinct magnetization states with different EHE Hall resistances and switching fields, therefore demonstrating a two- (or more) step magnetization reversal. Fig. 3 presents an experimental implementation of an eight-state memory cell composed of two films, the first exhibiting a single step reversal and the second exhibiting a double-step reversal. Samples 3 and 4 were fabricated by e-beam deposition of a Co/Pd multi-strata structure composed of 0.2 nm thick Co and 0.9 nm thick Pd strata repeated six times [Co 0.2 / Pd 0.9]$_6$ deposited on a 7.5 nm thick Pd seed layer. Sample 3, exhibiting a single step reversal, was deposited on a glass substrate and sample 4 demonstrating a clear double state magnetization reversal (Fig.3a open circles) was deposited on a GaAs substrate. The measurement was done at 77K. Fig.3b presents the inter-dot EHE voltage measured during a sequence of minor loops by using the RMFR protocol. Eight distinct remnant states corresponding to eight possible magnetization states in two films are obtained by following the corresponding field sequences. Three "writing" field magnitudes were used: $H_{w1}$, $H_{w2}$, $H_{w3}$ such that $H_{w1} > H_{s1} > H_{w2} > H_{s2} > H_{w3} > H_{s3}$, where $H_{s1}$, $H_{s2}$ and $H_{s3}$ are three switching fields corresponding to three storage layers in the cell.

Sixteen memory states are illustrated in Fig.4. The two films used in the cell are samples 4 and 5. Sample 5 was fabricated on a GaAs substrate by e-beam deposition of a two-level Co/Pd structure where the first component, composed of 0.2 nm thick Co and 0.9 nm thick Pd layers repeated six times [Co 0.2 / Pd 0.9]$_6$ was followed by 3 nm thick Pd spacer, followed by the second level composed of 0.4 nm thick Co and 0.9 nm thick Pd layers repeated six times [Co 0.4 / Pd 0.9]$_6$. The EHE signal of sample 5 is shown in Fig.4a by open circles. Notably, the unusual form of the EHE hysteresis of this sample is due to different polarities of the EHE effect in the two component layers: [Co 0.2/ Pd 0.9]$_6$ and [Co 0.4/ Pd 0.9]$_6$. Comprehensive understanding of the reversal of the EHE polarity with thickness is still lacking [12]. The EHE inter-dot voltage measured in the cell is shown in Fig.4b for an assembly of minor loops. Sixteen distinct remnant states are obtained at zero field corresponding to the respective field sequences.

The principle of a split cell composed of several dots can have different embodiments. Here we demonstrated a cell composed of two dots connected in series, while storage layers in each dot are connected in parallel. An alternative type of memory cell can contain N dots (N ≥ 2) connected in parallel. The number of possible memory states in such a cell is $2^{\sum_{i=1}^{N} n_i}$.

An important challenge in implementation of multi-bit magnetic memory is magnetic coupling among different dots and segments of multi-level structures. As has been shown by Albrect et al [3] in systems with perpendicular anisotropy, the different storage layers generate demagnetization fields that favor the ferromagnetically aligned states over the mixed ferrimagnetic states, which makes the states with antiparallel magnetization unstable. The problem is acute for fabrication of multiple storage layer media with thinner spacer layers which are required to reduce spacing loss in hard disk recording systems. In our case, distances between dots and spacing between different layers of the same multilevel dots can be sufficiently large to suppress coupling among individual storage layers. Therefore, in addition to multiplicity of states, probably the most important advantage of the split cell architecture is freedom in positioning dots of the same cell at separate locations and different levels, thus building an effective three-dimensional memory.

This work was supported by the Israel Science Foundation Grant No.633/06.

# References.

**Figure captions.**

Fig.1 Schematic presentation of a memory cell containing two ferromagnetic dots connected in series.

Fig.2. Realization of a four-state memory cell. (a) EHE hysteresis loops of sample 1 (full circles) and 2 (open circles). (b) EHE component of the inter-dot voltage obtained by following Eq. 5. Solid circles indicate the major hysteresis loop. Open circles indicate two minor loops at which the field sweep is reversed when magnetization vectors in two films are anti-parallel, marked by u-turn arrows. (c) The total inter-dot voltage $V_{AD,FE}$ measured during a sequence of field ramps. T ≈ 295 K. Field is applied perpendicular to plane.

Fig. 3. Realization of an eight-state memory cell. (a) EHE hysteresis loops of samples 3 (full circles) and 4 (empty circles). (b) EHE component of the inter-dot voltage obtained by following Eq. 5 for a series of different minor loops. T = 77 K. Field is applied perpendicular to plane.

Fig. 4. Realization of a sixteen-state memory cell. (a) EHE hysteresis loops of samples 4 (full circles) and 5 (empty circles). (b) EHE component of the inter-dot voltage obtained by following Eq. 5 for a series of different minor loops. T = 77K. Field is applied perpendicular to plane.

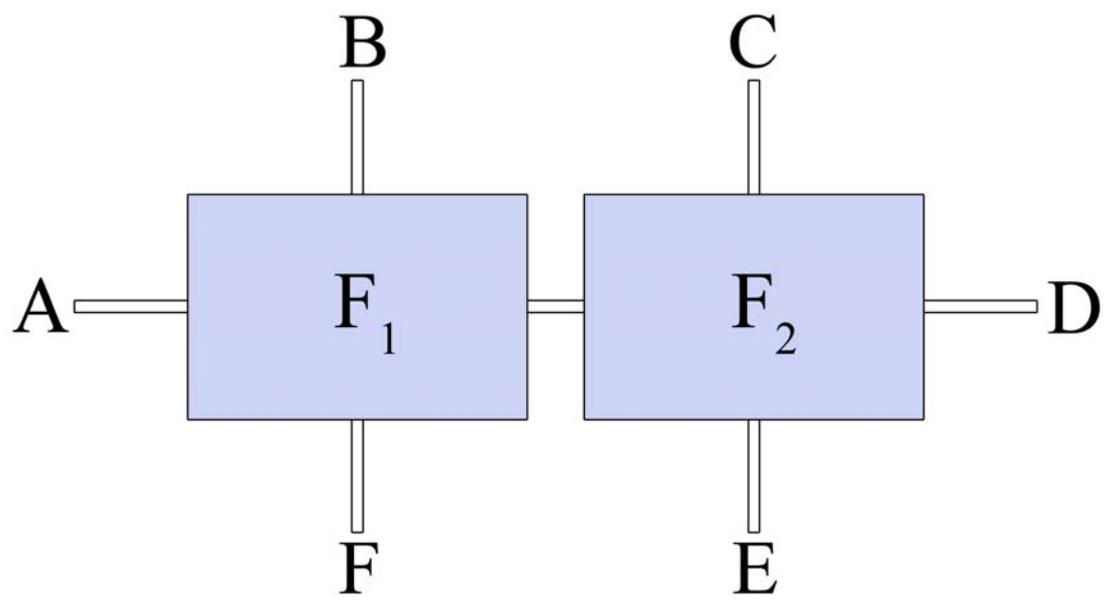

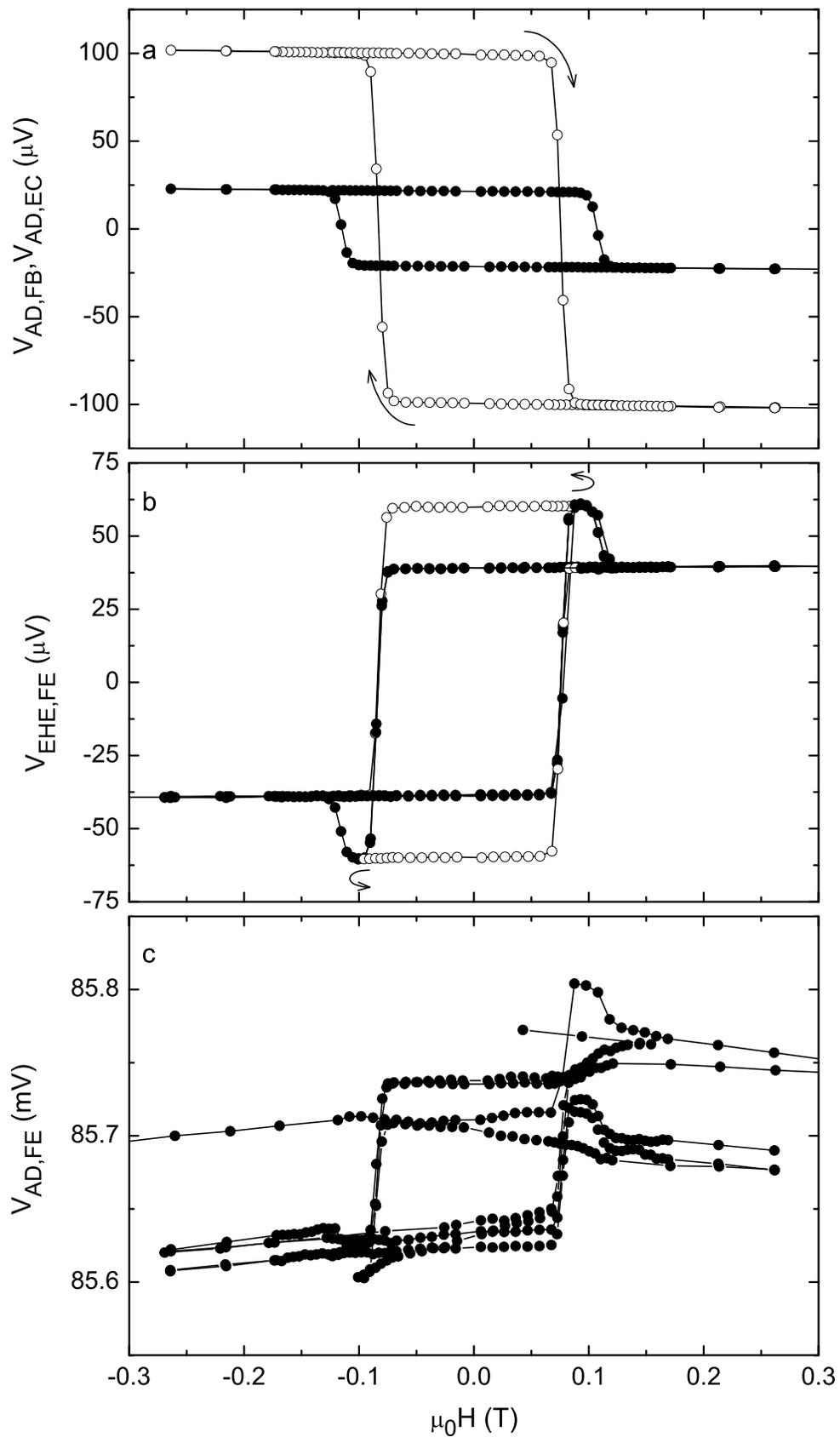

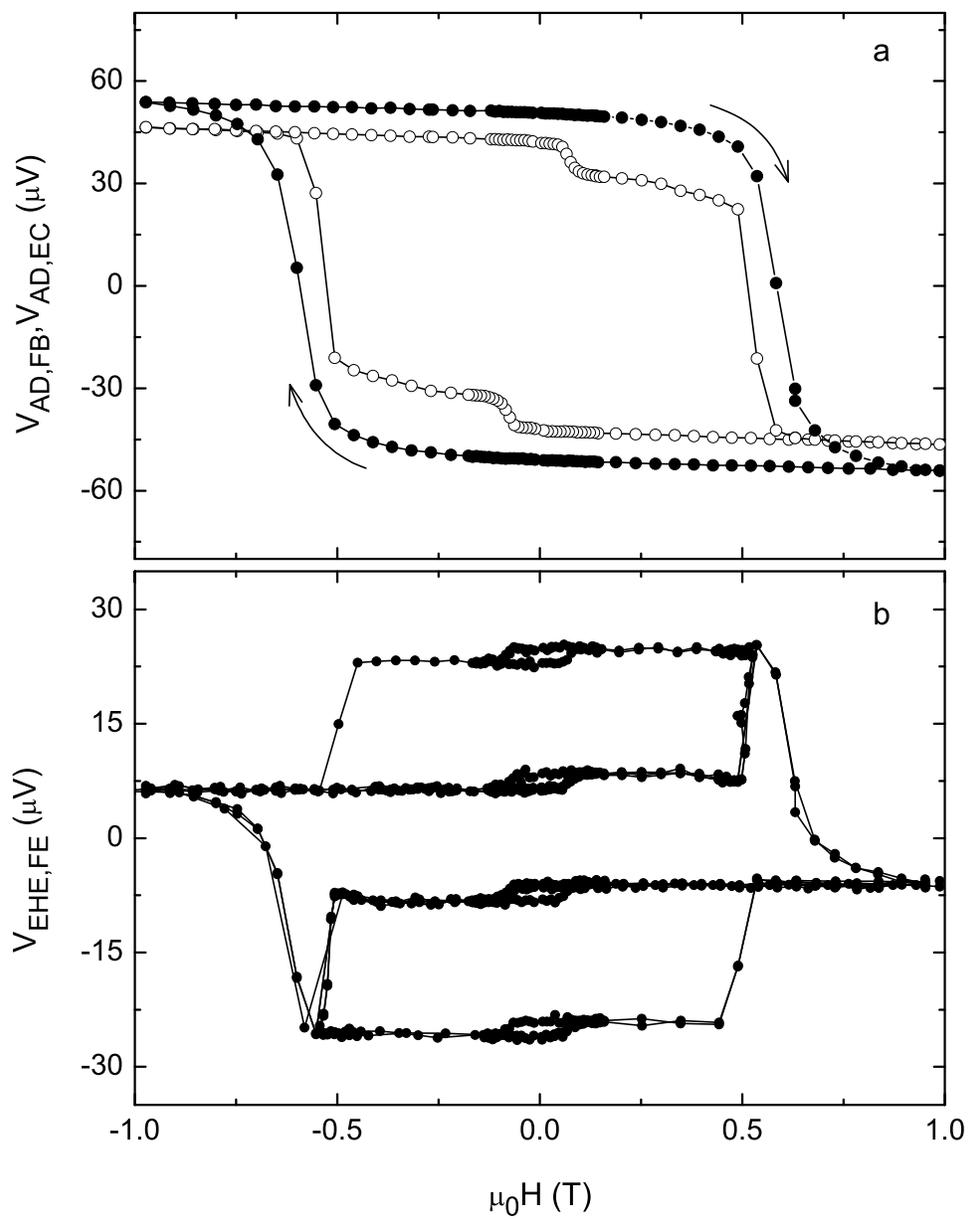

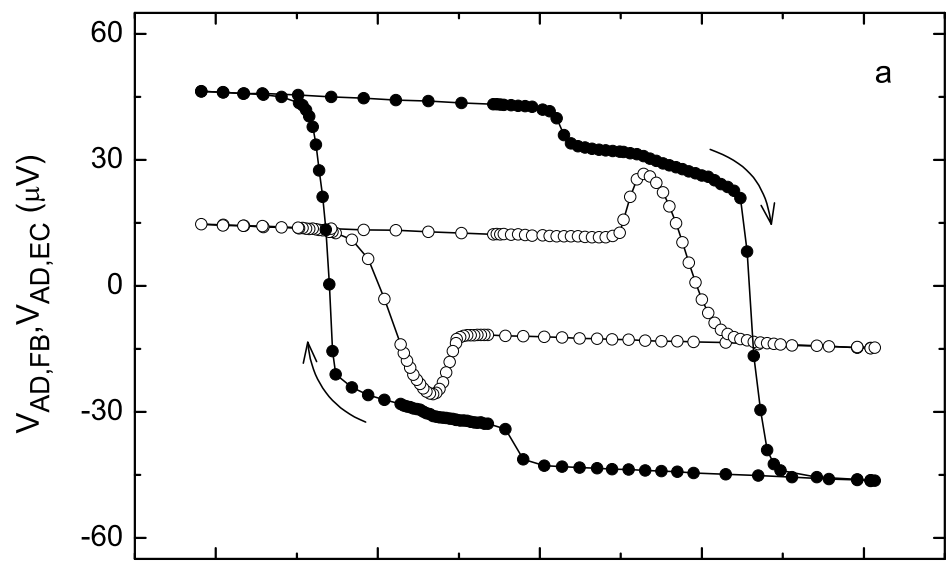
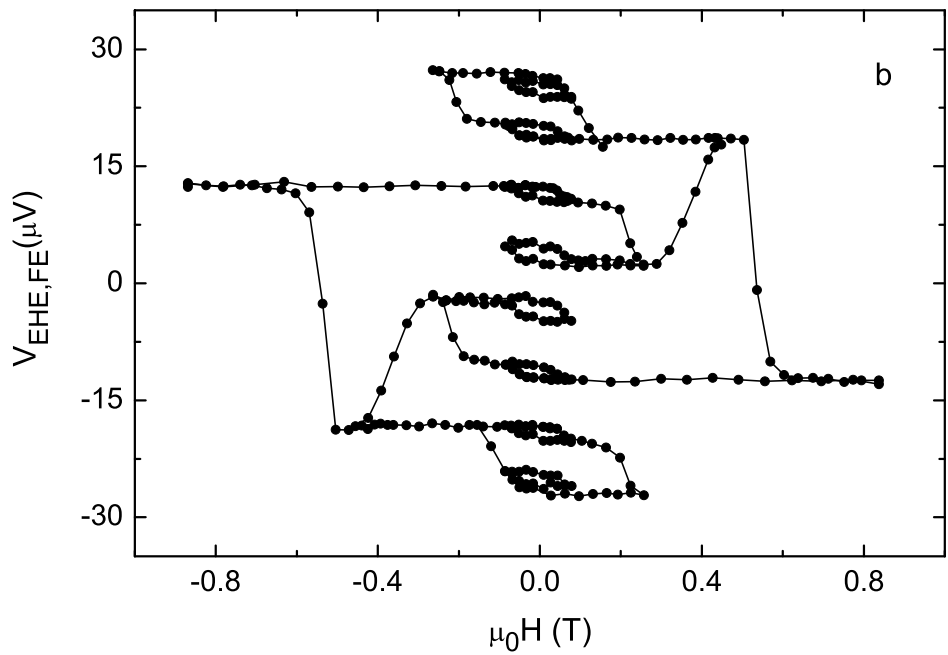